\documentclass[aps,prb,twocolumn,showpacs]{revtex4}
\usepackage{graphicx}

\begin{document}
\title{Non monotonic velocity dependence of atomic friction}

\author{Enzo Granato}
\address{Laborat\'orio Associado de Sensores e Materiais,
Instituto Nacional de Pesquisas Espaciais, \\
12227-010 S\~{a}o Jos\'e dos Campos, S\~ao Paulo, Brazil}
\author{S.C. Ying}
\address{Department of Physics,
Brown University, \\
Providence, Rhode Island 02912, USA}


\begin{abstract}

We study the velocity dependence of the frictional force of the tip
of an atomic force microscope as it is dragged across a surface,
taking into account memory effects and thermal fluctuations. Memory
effects are described by a coupling of the tip to low frequency
excitation modes of the surface in addition to the coupling to the
periodic corrugation potential. We find that when the excitation
mode frequency is comparable to the characteristic frequency
corresponding to the motion of the tip across the  surface, the
velocity dependence of the frictional force is non monotonic,
displaying a velocity range where the frictional force can decrease
with increasing velocity. These results provide theoretical support
for the interpretation of recent experiments which find a frictional
force that decreases with velocity on surfaces covered with a
monolayer.

\end{abstract}


\maketitle

\section{Introduction}
In an atomic  force microscope, the motion of the tip as it is
dragged across a substrate provides an efficient way to probe atomic
friction of surfaces. In fact, the moving tip can be regarded as a
single asperity which determines the frictional force between
macroscopic surfaces  \cite{Perssonbook}. Under a constant load, the
low-velocity motion of the tip on a surface exhibits stick-slip
behavior as the moving tip hops over the corrugation potential
defined by the substrate. For a quantitative description of friction
at the microscopic level, theoretical understanding of the behavior
of the tip on the substrate under different conditions is required.
Recent studies have argued that due to thermal fluctuations, the
lateral frictional force increases logarithmically with increasing
sliding velocity \cite{Gnecco,Grant,Riedo03}. However, it is also
found experimentally that the frictional force can have an opposite
behavior, where it decreases with the sliding velocity depending on
the ambient conditions \cite{Riedo02} and the nature of the
substrate \cite{Salmeron}. The origin of this decreasing velocity
dependence is not fully understood and it is possible that distinct
effects need to be taken into account to explain different
experimental conditions. Current explanations of this behavior for
point contact friction are based on extensions of the simplest
effective model, the generalized Tomlinson model
\cite{Grant,Urbakh,Muser}, taking into account additional effects
such as dissipation due to deformation of the AFM tip \cite{Reimann}
or the inclusion of additional time scales for reorganization of the
surface during sliding \cite{Salmeron}. In particular, the
interpretation of recent experiments  on surfaces covered with a
monolayer in terms of an additional time scale due to surface
restructuring \cite{Salmeron} still lacks detailed theoretical
support.

In this work, we study the velocity dependence of the frictional
force within a simple theoretical model taking into account the
coupling of the AFM tip to excitation modes of the surface in
addition to the coupling to the periodic corrugation potential. We
model the generic excitation modes by damped harmonic oscillators
analogous to the model of an adsorbate coupled to phonon modes of
the substrate \cite{Cuccetti}. We find that when the excitation mode
frequency is comparable to the characteristic frequency of the
particle moving across the surface potential, the velocity
dependence of the frictional force is non monotonic displaying a
velocity range where the frictional force can decrease with
increasing velocity. These results provide theoretical support
within a simpler model for the interpretation of recent experiments
\cite{Salmeron} which find a frictional force that decreases with
velocity on surfaces covered with a monolayer.

\section{Model}
The lateral motion of the tip can be described by a model of  a
single particle in an external one-dimensional potential
representing the substrate and coupled elastically to the moving
support. The equation of motion for the particle is given by
\cite{Grant}
\begin{equation}\label{tipeq}
m \ddot{x} = -\frac{ d V(R,x)}{d x} -m\gamma \dot{x}+ f(t)
\end{equation}
where the noise variables $f(t)$ satisfy
\begin{equation} \label{noise}
<f(t) f(t')> = 2 m \gamma k_B T \delta (t -t')
\end{equation}
and $V(R,x)=V_p(x) + V_s(x,R)$ is the total potential including the
interaction of the tip to the substrate with periodic potential
\begin{equation}
V_p(x) = V_o[1- cos(2 \pi \frac{x}{a})]
\end{equation}
and the elastic interaction between the tip and the moving support
\begin{equation}
V_s(x,R) = \frac{1}{2} k (x- R(t))^2
\end{equation}
where $R(t)$ is the position of the support.

The friction damping parameter $\gamma$ in Eq. (\ref{tipeq}) results
from the coupling of the tip to the surface excitations with a much
shorter time scale than the motion of the tip (e.g. electronic
excitations) which results in a $\delta$-function correlated random
force $f(t)$ with instantaneous Markovian damping without memory as
described by Eq. (\ref{noise}). The microscope support is moving at
constant velocity, $ R(t) =v t$, where $v$ is the support velocity.
The average force on the support due to the tip motion is then given
by
\begin{equation}\label{force}
F=k<R(t)-x>
\end{equation}
This is regarded  as the frictional force with the substrate since
without damping we should have $F=0$ when averaging out the negative
and positive force region of the periodic potential. For
sufficiently large frictional damping
stick and slip motion is expected. As shown in Ref.
\onlinecite{Grant}, using this model the frictional force $F$
increases with increasing velocity $v$. At small $v$, it increases
logarithmically but ultimately one would expect $F$ to be just
$\gamma v$, for sufficiently large $v$.

We extend the model described by Eq. (\ref{tipeq})  by
allowing the tip to couple to other excitations in the substrate
which have a finite time scale. These
other excitations are modeled by damped simple harmonic oscillators as done
in Ref. \onlinecite{Cuccetti} . These excitations can be
labeled by the mode index $\lambda=(i,l)$ where $i=1,2$ labels the
symmetry group and $l $ runs from $1$ to $N/2$ where $N$ is the
total number of modes. Each mode is characterized by a displacement
variable $u_\lambda$, which couples to the tip motion. The total
Hamiltonian can be written as
\begin{eqnarray}
H & = & \frac{p^2}{2m} + V(x,R) - \sum_\lambda\frac{M
\omega_\lambda^2}{2} W_\lambda^2(x) \cr &+&
\sum_\lambda[\frac{p_\lambda^2}{2M}+\frac{M\omega_\lambda^2}{2}(u_\lambda+W_\lambda(x))^2]
\end{eqnarray}

The third term in the above Hamiltonian represents a counterbalance
term. This is  added so that when the tip moves very slowly relative
to all the time scales given by $ 2 \pi / \omega_\lambda$, then at
any instantaneous position of $x$, $u_\lambda=-W_\lambda$ and the
actual effective potential felt by the tip is  $ V_p(x)- \sum
\frac{M \omega_\lambda^2}{2}W_\lambda^2 $ which is the adiabatic
potential. However, when the tip is moving very fast relative to
such time scales, then the substrate has no time to respond, and one
recovers the rigid substrate potential $V_p(x)$. The explicit form
of $W_\lambda (x)$ is chosen as
\begin{eqnarray} \label{coupl}
W_{l,l} & = & \frac{2 \pi \alpha V_o}{a \sqrt{\frac{N}{2}}M \omega_l^2 }
sin (2 \pi x / a) \cr W_{2,1} & = & \frac{2 \pi  \alpha V_o}{a
\sqrt{\frac{N}{2}}M \omega_l^2} cos (2 \pi x / a)
\end{eqnarray}
The coupling $\alpha$ in Eq. (\ref{coupl}) depends on the frequency
$\omega$ of the mode. This is chosen here to have the form
$\alpha(\omega) \propto \omega$ so that  the total effect of a
finite number of damped oscillators corresponds to a continum
distribution \cite{Cuccetti} of excitation modes with a  density of
states $\rho(\omega)\propto \omega^2$, analogous to that for 3-D
phonons in the Debye model. The equations of motion for the tip and
the mode displacement variables can be written in a dimensionless
form as
\begin{eqnarray} \label{dimeq}
\ddot x & =  &  - 2 \pi sin( 2 \pi x) + k(R(t) - x)  - 2 \pi
\alpha_o \sum_\lambda \omega_\lambda f_\lambda( x ) u_\lambda   \cr
& & - \gamma \dot x + f(t) \cr \ddot u_\lambda  & = &
\omega_\lambda^2 u_\lambda - \alpha_o \ \omega_\lambda g_\lambda(x)
- \gamma_\lambda \dot u_\lambda + r_\lambda(t)
\end{eqnarray}
where the additional noise variables $r_\lambda$  for the damped
harmonic oscillator also satisfy a fluctuation-dissipation relation
\begin{equation}
<r_\lambda(t),r_\lambda(t')> = 2 \gamma_\lambda  T \delta (t
-t') \delta_{\lambda, \lambda'}
\end{equation}
Without loss of generality, we can set $m=M$ for simplicity. In Eq.
(\ref{dimeq}), $f_{1,l}=cos(2\pi x)$, $g_{1,l}=sin(2 \pi x)$,
 $f_{2,l}=sin(2 \pi x)$, $g_{2,l}=cos(2 \pi x)$ and $\alpha_o=2 \pi
 V_o/(a \sqrt{N/2})$. The lattice constant $a$ is taken as the unit
 of length, $V_o$ the unit of energy, and $\sqrt{m a^2/V_o}$ the
 unit of time.


\section{Numerical results and discussion}

We have performed numerical simulations of the coupled dynamical
equations (\ref{dimeq}) using Brownian molecular dynamics
\cite{Allen}. The parameters for the spring and microscopic damping
were fixed to  $k=3$ and $\gamma=6$  and for damped harmonic oscillators,
$\gamma_\lambda = 2 w_\lambda$ and $\alpha_o=2.5$. Time steps ranged
from $dt=0.001$ to $0.005$. Numerical results for the velocity
dependence of the friction force obtained from Eq. (\ref{force})
assuming different angular frequencies $\omega$ for coupling to one odd
or even mode are shown in Figs 1 and 2. They are qualitative
similar and show that when the frequency of the excitation mode is
sufficiently small the velocity dependence of the friction force is
non monotonic. While at sufficiently small or large velocities  it
increases with velocity there is an intermediate range of velocities
where the frictional force decreases as the velocity increases. The
approximate linear decrease in the semilog plot of Figs 1 and 2,
suggest that this decrease is logarithmic.

\begin{figure}
\includegraphics[ bb= 2cm 3cm 18cm 17cm, width=7.5 cm]{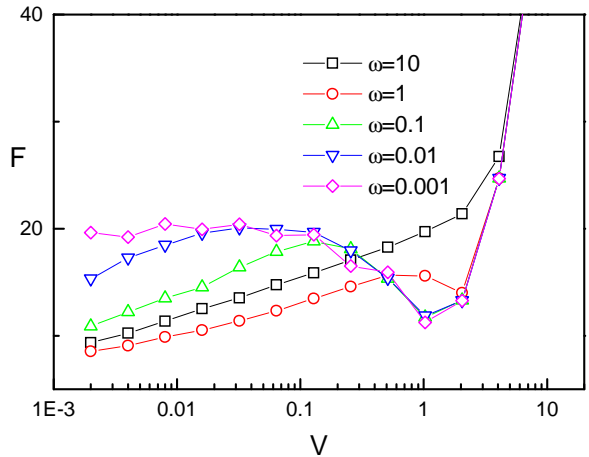}
\caption{ Velocity dependence of the frictional force of the tip
moving on a substrate coupled to an excitation mode with different angular
frequencies $\omega$. Temperature $ kT /V_o= 0.3$. Results are for
the odd-mode model.}\label{odd}
\end{figure}

\begin{figure}
\includegraphics[ bb= 2cm 3cm 18cm 17cm, width=7.5 cm]{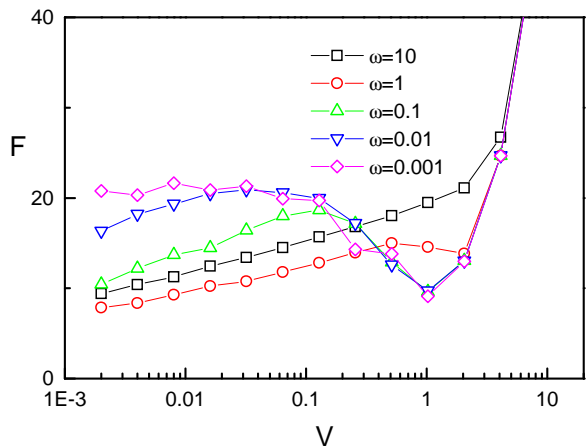}
\caption{ Velocity dependence of the frictional force of the tip
moving on a substrate coupled to excitation mode   with different angular
frequencies $\omega$. Temperature $ kT /V_o= 0.3$. Results are for
the even-mode model.}\label{even}
\end{figure}

To understand this non monotonic behavior it should be noted that
there are several important time scales in the problem. First, there
is a velocity dependent time scale, $t_v = a/v$, which is the
average time for the tip to traverse one lattice period of pinning
potential. The pinning potential introduces a time scale, $t_0=
\sqrt{m a^2 /V_o}$, the vibrational period of the tip in the well.
In the stick and slip regime where the tip is stuck in the well and
perform many vibrational oscillations before hopping to the next
minima, we have $t_v >> t_0$. The time for crossing the barrier is
relatively fast, on the order of $t_0$. With the coupling to a new
excitation mode of frequency $\omega_l$, another  time scale $t_l =
2 \pi/\omega_l$ is introduced. We now consider what is the effect of
varying $t_v = a/v$ on the frictional force $F$. In the absence of
the coupling to the  excitation  mode, the logarithmically increase
of the friction force with velocity \cite{Grant} was obtained for
velocities up to $ t_v \approx t_0$. Now consider what happen if we
choose an excitation mode such that $t_l$ is several times larger
than $t_0$ ($ \omega_l << \omega_0$ ) and we vary $t_v$. As far as
the intrinsic frictional damping is concerned, when $t_v$ approaches
$t_l$, the damping should increase and also becomes non-Markovian.
Eventually, when $t_v$ becomes very large, the damping should
saturate to a Markovian value. So based on this consideration, the
effective $\gamma$ should increase and saturate as a function of
decreasing $v$. The frictional force $F$
should exhibit similar behavior.  
In addition, the even mode can lead to a velocity dependent
substrate relaxation that changes the effective potential as
discussed previously. However, for the present choice of parameters,
the velocity dependence of the effective non-adiabatic frictional
force due to the coupling to the new excitation mode dominates as
evidenced by the results in Fig. 1 and 2, showing similar results
for coupling to an  even or odd modes.

\begin{figure}
\includegraphics[ bb= 2cm 3cm 18cm 17cm, width=7.5 cm]{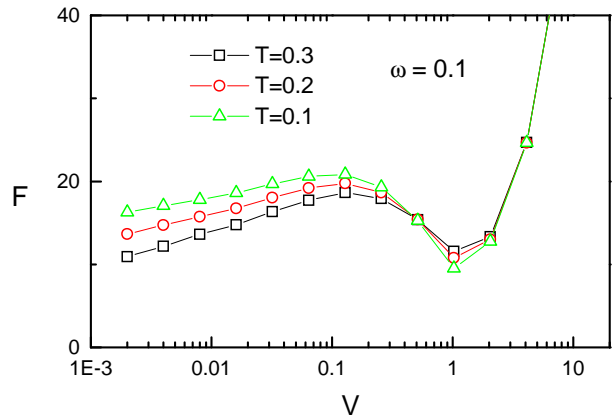}
\caption{ Velocity dependence of the frictional force of the tip
moving on a substrate coupled to an excitation mode of angular frequency
$\omega =0.1$ at different temperatures $T/V_0$. Results are for the
odd-mode model.}\label{todd}
\end{figure}

The effect of varying the temperature at fixed mode  frequency $\omega$ is
also of interest. As can been from Fig. 3, the qualitative non
monotonic behavior remains but there is an increase in the magnitude
of the negative slope of the velocity-force curve as the temperature
decreases.

We have also performed additional calculations to check the effect
of including more excitation modes to model a more realistic
continuum model of substrate excitations. $6$ independent damped
harmonic oscillators with frequencies in a range $[\omega_{min},
\omega_{max}]$ with $\omega_{min}=\omega_{max}/6$ were included in
these calculations. As can been seen from Fig. 4, the behavior for
the velocity dependence for different maximum frequencies
$\omega_{max}$ is qualitative similar to the behavior in Fig. 1 at
the same temperature. The main effect of the additional modes below
$\omega_{max}$ is a tendency to saturation of the frictional force
at lower frequencies.

\begin{figure}
\includegraphics[ bb= 2cm 3cm 18cm 16cm, width=7.5 cm]{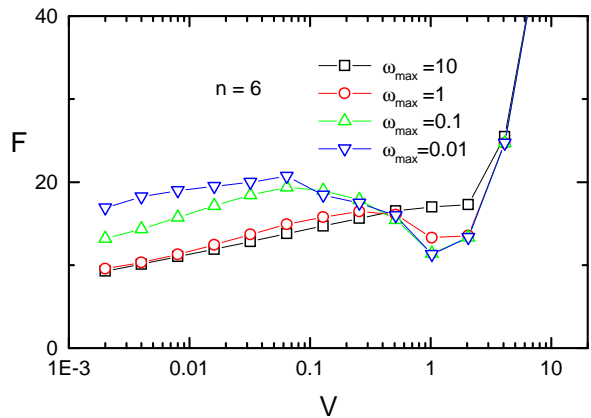}
\caption{ Velocity dependence of the frictional force of the tip
moving on a substrate with excitation modes ranging from
$\omega_{min}$ to $\omega_{max}$. $6$ modes  equally spaced with
$\omega_{min}=\omega_{max}/6$ are used. Temperature $ kT /V_o= 0.3$.
Results are for the odd-mode model..}\label{band}
\end{figure}

\section{Conclusion}

We study the velocity dependence of the frictional force of the tip
of an atomic force microscope as it is dragged across a surface
using a single particle model, taking into account memory effects
and thermal fluctuations. In our model, there is  an additional
coupling of the particle to excitation modes of the surface modeled
by damped harmonic oscillators \cite{Cuccetti}. We find that when
the excitation  mode frequency is comparable to the characteristic
frequency of the motion of the tip across the surface potential, the
velocity dependence of the frictional force is non monotonic
displaying an intermediate velocity range where the frictional force
can decrease with increasing velocity. Therefore, the sliding
behavior on the surface depends on the nature the surface through
the characteristic frequency of such excitation modes. Recently,
non-monotonic velocity dependence of friction was observed
experimentally \cite{Salmeron} on surfaces covered with a monolayer.
The experimental results were interpreted as resulting from the
chemical nature of the surface, with friction increasing or
decreasing with velocity depending on the presence of cross-linked
H-bonds. It was argued that for higher sliding velocity there is not
enough time for reordering to occur and the friction force should
decrease logarithmically with velocity, while at lower velocities
the breaking and reordering of the H-bonds contribute a new source
of friction that increases with decreasing velocity. Our results
provide theoretical  support for this idea through an explicit
calculation within a simple model. The breaking of glassy domains of
H-bonds at a critical stress and reordering introduces a new time
scale, which corresponds in our model to the inverse frequency of
the additional excitation mode of the surface. The non-monotonic
velocity dependence of the friction then follows naturally when the
time scale of the motion of the tip across the surface becomes
comparable to this new time scale for the excitation mode. However,
since we find a non monotonic behavior without invoking any glassy
behavior, our results suggest that similar behavior should be found
even on surfaces without disorder when additional excitation modes
are present.

\section{Acknowledgments}

We thank T. Ala-Nissila for helpful discussions and suggestions.
E.G. was supported by Funda\c c\~ao de Amparo \`a Pesquisa do Estado
de S\~ao Paulo - FAPESP (Grant No. 07/08492-9). S.C.Y. also
acknowledges FAPESP (Grant No. 09/01942-4) for supporting a visit to
Instituto Nacional de Pesquisas Espaciais.

\end{document}